\documentclass[conference]{IEEEtran}

\usepackage{float}
\usepackage{mdwmath}
\usepackage{mdwtab}
\usepackage{amsfonts}
\usepackage{algorithm}
\usepackage{algpseudocode}
\usepackage{lipsum}
\usepackage{cite}
\usepackage{algorithmicx}
\usepackage{algpseudocode}

\ifCLASSINFOpdf
  \usepackage[pdftex]{graphicx}

\else

\fi

\usepackage[cmex10]{amsmath}

\usepackage{mdwmath}
\usepackage{mdwtab}

\usepackage{graphicx}
\usepackage{algorithm}
\usepackage{algpseudocode}
\usepackage{lipsum}
\usepackage[cmex10]{amsmath}
\usepackage{amsfonts}
\usepackage{verbatim}
\usepackage{verbatim}
\usepackage{algpseudocode}
\usepackage{multicol}
\usepackage{filecontents}

\newcommand{\qed}{\nobreak \ifvmode \relax \else
      \ifdim\lastskip<1.5em \hskip-\lastskip
      \hskip1.5em plus0em minus0.5em \fi \nobreak
      \vrule height0.75em width0.5em depth0.25em\fi}

 \usepackage{tikz}
        \usepackage{tikz-qtree}
        \usetikzlibrary{matrix}

\usepackage{subcaption}
\usepackage{kantlipsum} 
\usepackage{mwe} 
\usepackage{tablefootnote}
\usepackage{caption}
\DeclareCaptionLabelSeparator{periodspace}{.\quad}
\begin{document}

\title{Performance Evaluation of VDI Environment}
     
\author{\IEEEauthorblockN{
Hafiz~Ur~Rahman\IEEEauthorrefmark{1},
Farag~Azzedin\IEEEauthorrefmark{1},
Ahmad~Shawahna\IEEEauthorrefmark{2},
Faisal~Sajjad\IEEEauthorrefmark{1}, and
Alyahya~Abdulrahman\IEEEauthorrefmark{1}
}
\IEEEauthorblockA{\IEEEauthorrefmark{1}Information and Computer Science}
\IEEEauthorblockA{\IEEEauthorrefmark{2}Department of Computer Engineering}

King Fahd University of Petroleum and Minerals,
Dhahran-31261, KSA\\ 

\{g201404780, fazzedin, g201206920, g201409220, g201101050\}@kfupm.edu.sa
}

\maketitle


\begin{abstract}
Virtualization technology is widely used for sharing the abilities of computer systems by splitting the resources among various virtual PCs. In traditional computer systems, the utilization of hardware is not maximized and the resources are not fully consumed. By using the virtualization technology, the maximum utilization of computer system hardware is possible. Today, many organizations especially educational institutes are adopting virtualization technology to minimize the costs and increase the flexibility, responsiveness, and efficiency. In this paper, we built a virtual educational lab in which we successfully implemented Virtual Desktop Infrastructure (VDI) and created more than 30 virtual desktops. The virtual desktops are created on the Citrix XenServer hypervisor using the Citrix XenDesktop. Finally, we used benchmarking software tool called Login VSI in order to measure the performance of the virtual desktop infrastructure and the XenServer hypervisor. The results have shown that the system performs well with 30 virtual desktops without reaching the saturation point.

\end{abstract}


\IEEEpeerreviewmaketitle


\section{Introduction}
The word computing in computer science is renovated into a model that consists of different services which are used to be delivered to the end user’s in a similar way that we delivered the traditional utilities such as telephony, water, and electricity services \cite{buyya2009cloud}. In such model, the end users have no concerns about how the services are delivered and where they are hosted. Several computing patterns are used to deliver these utilities such as grid computing, cluster computing, and cloud computing. The recent progress in computing vision called cloud computing which provides a convincing value intention for organizations to contract out their Information and Communication Technology (ICT) infrastructure \cite{haynie2009enterprise}. The cloud computing model allows the users and businesses to access their data and applications from any device and anywhere in the world whenever they want.

The cloud computing model provides millions of developing software on cloud to be consumed as a service, this solves the problem for individual computers where they run these software. The cloud infrastructure is monitored and maintained by the different content providers such as Amazon, Sun Microsystem, Google, Microsoft, and IBM. Before the advent of cloud computing, the traditional data centers were used to provide the IT services and application to the end users. However, the traditional data centers have several shortcomings such as the data/information cannot be updated in real time, the maintenance is costly, sufficient effort and much time is required in order to maintain the data centers. Moreover, sometimes the technical people are required to manage and control the traditional data center systems. In addition, one of the big disadvantages of data centers is inflexibility. Therefore, we cannot move and exchange the services among the physical servers. Some organizations have limited budget, they cannot afford the cost of data centers since the data centers have to be kept for a long time and need to be upgraded which is considered time consuming and costly. Therefore, cloud computing is cheaper for the organizations that cannot afford traditional data centers since the cloud computing is less costly \cite{nakazawa2012influence, paventhan2014towards}. Moreover, the data is more secured and updated in real time as compared to traditional data centers. The data in the cloud model is centralized in the data centers. The cloud computing model provides more features such as flexibility, scalability, performance, and space saving.

The cloud computing is very useful in educational process since the educational institutes lack of the resources such as providing computational software. The licensing of software for every student is a challenging for institutes. In educational institutes, every student and faculty member is required a physical machine in order to install and use the desired software. However, the institutes have lack of resources in which they cannot give physical system to every student in the institute. Therefore, this concludes a time consuming and costly process. The solution for this problem is virtualization.

The virtualization makes use of server resources in well-organized manner by setting up different types of servers within different cloud platforms such as public cloud and private cloud. It is very useful in education by providing easy access to the required resources via remote learning. The students and the faculty members can use computational software concurrently and efficiently. In addition, the virtualization helps in reducing the total cost of extra resources such as assigning each physical machine to student or faculty also it cuts the cost of licensing computational software. By using virtualization technology, we need to install the software on single machine while students and faculty can assess virtually. The virtualization has three types; server virtualization, desktop virtualization, and storage virtualization. In this paper, we used desktop virtualization.

\begin{figure*}
	\centering           
	\includegraphics[height = 7.5 cm, width=0.8\textwidth]{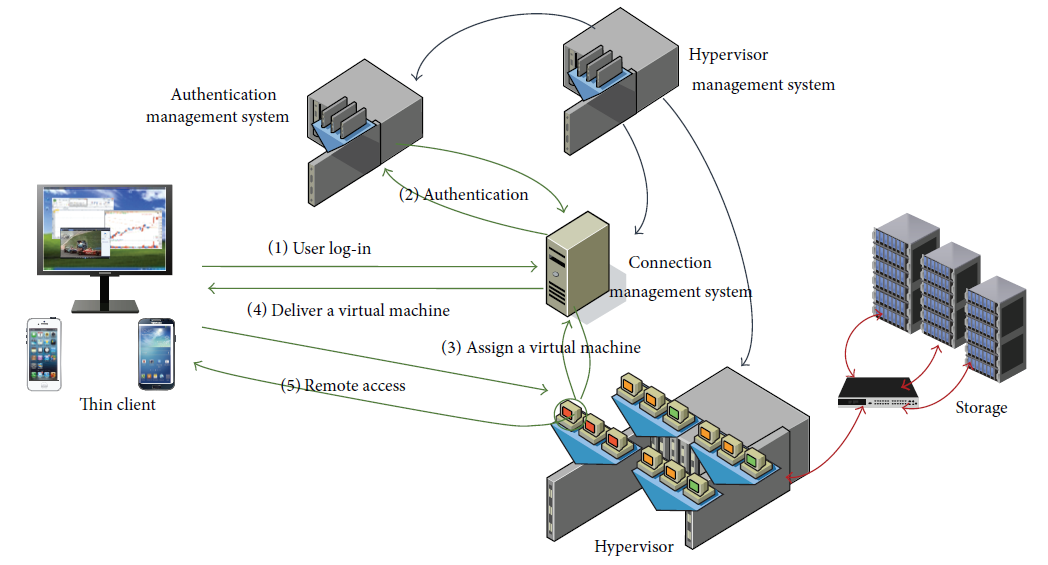}   
	\captionsetup{justification=centering}
	\caption{Hypervisor and VDI structure \cite{jeong2015investigation}.}  
	\label{fig1}       
\end{figure*}

The desktop virtualization is also called Virtual Desktop Infrastructure (VDI). It creates multiple logical desktops from a single physical machine. The VDI replaces the desktop PC by associating PCs as a Virtual Desktop (VD) \cite{dasilva2012enabling}. By using this technology, we can manage many virtual desktop from a single physical machine. The student and faculty can use their workspace using virtual desktop anytime and anywhere in the world through any workstation. Moreover, if the student or the faculty workstation is failed, their workspace will be safe and can be accessed through any other workstations. VMware, Citrix and Microsoft are widely used desktop virtualization solutions.

\begin{figure}[b]
	\centering           
	\includegraphics[height = 4.14 cm,width=0.5\textwidth]{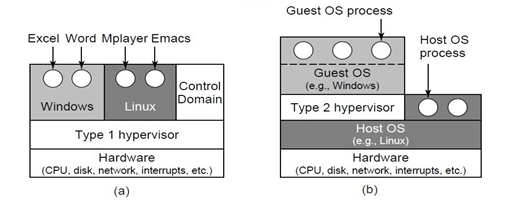}   
	\captionsetup{justification=centering}
	\caption{(a) Type 1 Hypervisor, and (b) Type 2 Hypervisor \cite{tanenbaum2014modern}.}  
	\label{fig2}       
\end{figure}

Figure~\ref{fig1} shows the hypervisor and VDI structure, where each user requests a login connection for virtual desktop. The user request is proceed to Connection Management System (CMS). The CMS take the login request information and pass it to the Authentication Management System (AMS). The CMS and AMS are very important for this whole process because they are involved in validating the user authentication and delivering the virtual desktop to the requested user. Once the user gets successfully authenticated from the AMS, the CMS initiates a request to the hypervisor to allocate virtual desktop for the requested user. When the virtual desktop is assigned to the requested user from the hypervisor, the CMS immediately delivers that virtual desktop to the requested user. Subsequently, the user can use the virtual desktop as it is a personal desktop. The storage devices are used for storing the Virtual Machines (VMs). The roles of Hypervisor Management System (HMS) are used to manage the hypervisor.

The hypervisor is responsible for creating different VMs, where each VM is used as virtual desktop. The resources such as memory, operating system, CPU, network, and data are shared among the different virtual desktops and completely managed by the hypervisor. Hypervisors are classified into two types; Type 1 hypervisor, and Type 2 hypervisor \cite{pawarperformance}. Figure~\ref{fig2} shows the structure of both hypervisors.

The main contribution of this paper includes studying the different types of hypervisors and VDI platforms, implementing more than 30 virtual desktops on a single physical machine, measuring the performance of Citrix XenServer hypervisor and virtual desktops using Login VSI benchmarking tool \cite{van2013virtual, vsiindexer}, and checking at which point the system performance will start decreasing.

The paper is organized as follows. Section II provides an extensive literature review of VDI. We discussed the methodology used to evaluate the VDI in section III. Experimental results and their analysis are presented in section IV. Finally, section V includes the conclusion and the future directions for this work.

\section{Literature Review}

Miseviciene et al. \cite{tuminauskas2012educational} discussed the challenges in the educational institutions such as the management of the data security and the cuts in budgets. The traditional IT resources of the educational institutions will become outdated after a while due to the inability to upgrade them in time. The authors have mentioned that the VDI technology will provide an innovation solution for these challenges. In addition, it will allow the current educational institutions for the remote learning integration at home. The authors applied the virtualization technologies in the cloud system at Kaunas University of Technology (KTU) in order to achieve the self-working at home. In such away, the university staff as well as the students will be able to access the university applications and virtual resources via Web. The authors have used Microsoft VDI platform with Hyper-V hypervisor for this purpose. Their experiment is considered homogeneous. The results have shown that a slightly noticeable increase in the uplink capacity has been monitored in the daily load.

The VDI technology has been introduced to the e-learning system due to its features such as flexibility and availability. However, the VDI is a network-based technology and therefore its performance will be subject to the bandwidth of the network used. The impact of the Quality of Service (QoS) on the e-learning environment arising from their use by the VDI has been investigated using the network emulator \cite{nakazawa2012influence, hirasawa2014learning}. The authors have used a XenServer hypervisor and a XenDesktop VDI platform for their experiments. The accomplished experiments are considered homogeneous since the used hypervisor and VDI platform are compatible and coming from the same vender. The results have clearly shown that it is suitable to use VDI in order to perform all workings acts for e-learning environment.

DaSilva et al. \cite{dasilva2012enabling} have discussed the reduction in IT budgets using virtual desktop infrastructure. There are high-rise in costs associated with current IT infrastructures of the traditional data centers. The current traditional IT infrastructure requires much electricity power to operate its wide machines and devices which leads to rising in the power bills. The authors have implemented a small virtual desktop environment using VMware and Wyse technology. They have used a vSphere hypervisor and a Horizon VDI platform from a VMware vender for their experiments. The results have shown that CPU and RAM have a direct influence on the power consumption of desktop PCs. In addition, they have shown that the shift to virtual desktop infrastructure offers a much more effective use of resources at the server level as well as the reduction of expenses for educational institutions.

Calyam et al. \cite{calyam2014vdpilot} discussed the challenges for faculty and students when accessing, reserving, and using the resources of universities labs. They implemented a VDI that can support up to concurrently 50 faculty and students access to the software available on the lab remotely over the Internet. The results have indicated that over 50\% of the participants found the virtual desktop user Quality of Experience (QoE) to be comparable to their home computer’s user QoE, and 8\% found the virtual desktop user QoE to be better than their home computer user QoE.

One of the main reasons that educational organizations has been attracted towards cloud computing is the sharp reduction of expenses \cite{baev2011cost}. Money can be saved by the less consumption of electricity power that cloud computing technologies can offer. In cloud computing, all educational services are residing on servers and centrally administrated. As a result, virtual labs can be easily implemented and deployed for students and instructors. In such away, the educational environment will always be ready and faster than the educational traditional environment. The authors have used a XenServer hypervisor and a XenDesktop VDI platform for their experiments. They implemented a VDI that can support up to 25 concurrent users.

The problems of the information structure in the university library system have been discussed in \cite{chen2014desktop}. The author focused on constructing and designing of VDI to allow the students to use the e-reading sources whenever they are. VDI can save time and effort for the IT technical team since they will be able to deploy many virtual desktops in a very short time. The author mentioned that maintaining of 120 virtual desktops takes less than 40 minutes. The author has used a vSphere hypervisor and a Horizon VDI platform from a VMware vender for the considered experiments. Also, the experiments accomplished are all considered homogeneous.

Paventhan et al. \cite{paventhan2014towards} have debated the deployment of cloud computing to extend the education services in order to achieve its benefits in schools. They mentioned that the cloud computing model is able to decrease cost and effort sharply when replacing IT computing resources. Also, the elasticity of provisioning cloud computing services is achieved to meet the dynamic demands and to utilize the cloud resources efficiently. In addition, the transition from the current traditional IT infrastructure to the cloud computing infrastructure is in general easy. Thus, cloud computing can be utilized in teaching and research for allowing contents of various courses and computing resources to be available all the time for students and faculty members, and to be easily remotely accessed from either on-campus or off-campus when it adopted by educational institutions. The authors have implemented well-known VDI platforms in their experiments such as Citrix XenDesktop, VMware Horizon, and Microsoft VDI. However, they did not evaluate these VDI platforms, and they did not specify the hypervisors used for their experiments.

\section{Research Methodology}

A systematic procedure has been adopted in this paper to measure the performance of virtual desktops and Type 1 hypervisor.

\subsection{Machines Setup}

In the first phase, two servers; control domain server, and infrastructure server were installed. The hardware specifications of both servers are almost same except the main memory and the hard drive. The hardware specifications of both servers are listed in Table~\ref{tab_1}.

\begin{table}[h]
	\renewcommand{\arraystretch}{1.2}
	\centering
	\captionsetup{justification=centering}
	\caption{Hardware specification of control domain server and infrastructure server.}
	\begin{tabular}{|c|c|c|}
		\hline
		Specification & Control Domain Server & Infrastructure Server\\
		\hline
		\hline
		Hardware Model & Intel Xeon & Intel Xeon\\
		\hline
		Processor Speed & 2 GHz & 2 GHz\\
		\hline
		CPU Processor & 12 Cores & 12 Cores\\
		\hline
		Logical Processors & 24 Cores & 24 Cores\\
		\hline
		Main Memory & 32 GB & 64 GB\\
		\hline
		Hard Drive  & 280 GB & 1107 GB\\
		\hline
	\end{tabular}
	\label{tab_1}
\end{table}

The control domain server has three VMs installed. The first VM is the active directory machine which is also called as AMS as discussed in Figure~\ref{fig1}, It is used to keep the records of users and authenticate the user credential whenever a user requests the virtual desktop. In addition, it provides the network infrastructure for the virtual desktops. The other two machines are used for the performance tools.

The infrastructure server is specifically selected for the VDI environment with two VMs and hypervisor. Citrix XenServer is used as a delivery controller for the virtual desktops. On the other hand, Citrix XenDesktop is used to create and manage the virtual desktops on hypervisor. The last virtual machine is the guest operating system in which it will be used for creating more than 30 virtual desktops for the users. However, these virtual desktops are stored in the storage devices as mention in Figure~\ref{fig1}. Note that each virtual desktop has one vCPU, 2 GB as the minimum requirement of RAM, and 24 GB for the disk drive.

\begin{figure}[b]
	\centering           
	\includegraphics[height = 5.75 cm, width=0.49\textwidth]{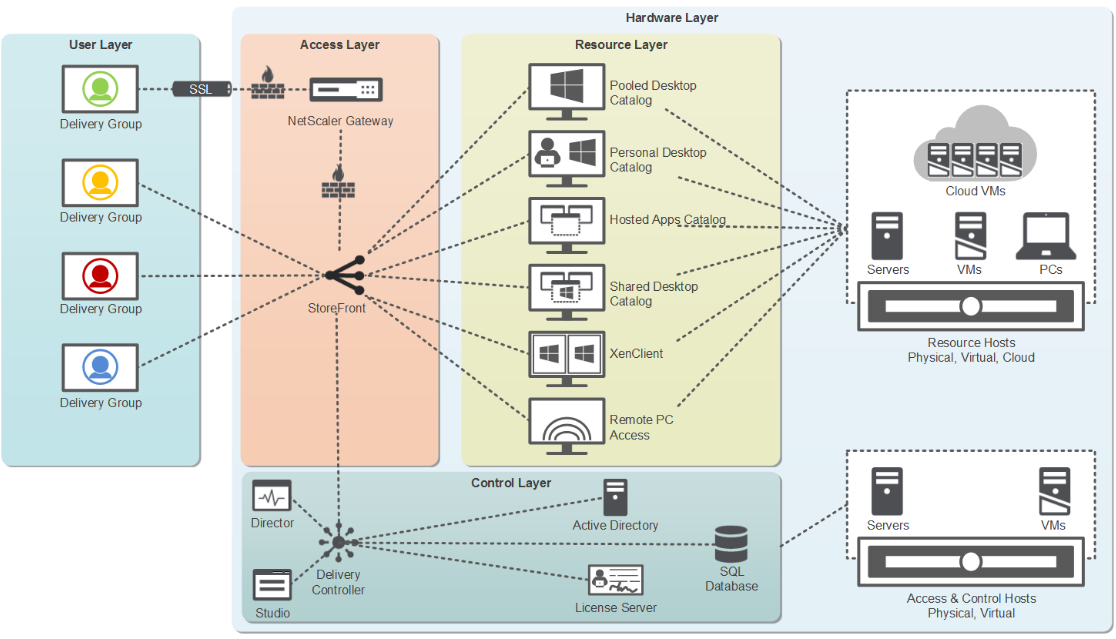}   
	\captionsetup{justification=centering}
	\caption{Conceptual diagram of Citrix XenDesktop \cite{silvestri2015citrix}.}  
	\label{fig3}       
\end{figure}

\begin{figure}
	\centering           
	\includegraphics[height = 5.20 cm, width=0.49\textwidth]{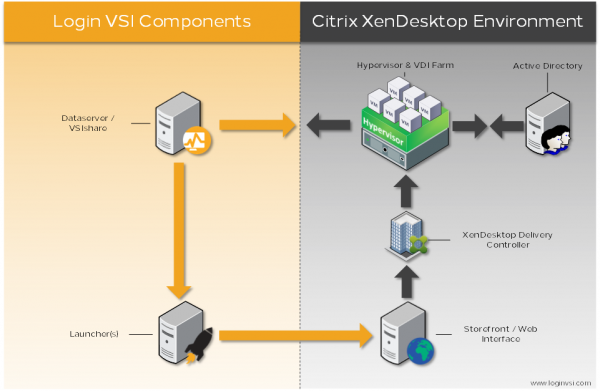}   
	\captionsetup{justification=centering}
	\caption{Conceptual diagram of Login VSI and components and Citrix XenDesktop \cite{van2013virtual}.} 
	\label{fig4}       
\end{figure}

Citrix XenServer is Type 1 hypervisor and one of the top open source virtualization platforms. Citrix XenServer does not require any host operating system; it works as hypervisor and host OS simultaneously in a single layer. Due to its abstraction, it increases the storage and server utilization in efficient way. In addition, Citrix XenServer gives the ability to create unlimited VMs and servers from a single console. Moreover, it is used to reduce the cost of power, physical storage space, and other equipment’s. Therefore, it has been selected in this research. In order to create VMs or virtual desktops on Citrix XenServer hypervisor, we need a management console called Citrix XenDesktop.

Citrix XenDesktop is the desktop virtualization solution from Citrix that enables virtual desktops as well as delivers applications to end users \cite{nakazawa2012influence, silvestri2015citrix}. Figure~\ref{fig3} provides the conceptual diagram of Citrix XenDesktop. The Citrix XenDesktop has several components such as Citrix Receiver, Delivery Controller, Virtual Delivery Agent, etc. These components are used for delivering and managing the virtual desktops.

\subsection{Performance Tool}

In order to measure the performance of VDI and hypervisor, Login VSI benchmarking tool \cite{van2013virtual} was used. This tool is installed on control domain server and it is responsible for performing various tests on the infrastructure server. The Login VSI was used in this work since it is one of the most popular tools that have been used in large scale industries where hundreds or thousands of virtual desktops are created. The Login VSI runs specific applications on each virtual desktop in order to check the performance of the hypervisor and VDI. In addition, it gives the ability to select the specification of user’s workloads to measure the scalability of virtual desktops. These workloads are simulated in virtual desktops to check the extreme capacity of load that the hypervisor can take. Figure~\ref{fig4} illustrates how Login VSI simulates the workload and tests the performance of VDI. The Login VSI has two components, namely, Launcher and Dataserver. The Launcher generates the workload on the target environment using the virtual desktop that is available in the storefront. The Dataserver is responsible for measuring the performance of the infrastructure.

\section{Results and Analysis}

In this section, the results that were obtained using the Login VSI tool have been discussed with the analysis for each of them. Each experiment was repeated five times and the results are averaged. The Login VSI tool provides four types of workloads for initiating the experiment; Test, Office, Knowledge, and Power workloads \cite{van2013virtual, vsiindexer}. The performance of the system was compared when using Office and Power workloads which are considered heavy and medium workloads, respectively.

\begin{figure}
	\centering           
	\includegraphics[height = 5.20 cm, width=0.49\textwidth]{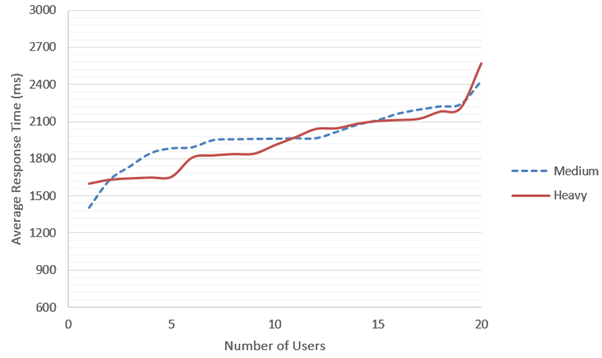}   
	\captionsetup{justification=centering}
	\caption{Average response time of Citrix XenDesktop and Citrix XenServer\\(20 users).} 
	\label{fig5}       
\end{figure}

\begin{figure}
	\centering           
	\includegraphics[height = 5.20 cm, width=0.49\textwidth]{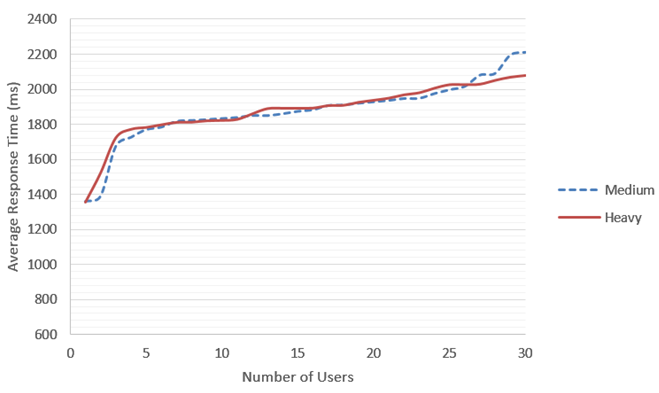}   
	\captionsetup{justification=centering}
	\caption{Average response time of Citrix XenDesktop and Citrix XenServer\\(30 users).} 
	\label{fig6}       
\end{figure}

Diverse applications are used to test the performance of virtual desktops and hypervisor. These applications are Microsoft Outlook, Microsoft Excel, Microsoft PowerPoint, Internet Explorer, 7-Zip, and Notepad \cite{van2013virtual}. Microsoft Outlook is used for sending messages, while Microsoft Word is used for typing documents. On the other hand, Microsoft Excel is used for generating large random data spreadsheets. Subsequently, the generated data are given to the CPU for computation and written to the disk using I/O operations. Microsoft PowerPoint is used for editing and reviewing the slides. However, Internet Explorer is used for browsing different websites. On the other hand, 7-Zip tool is used for compressing data with low and high compression rate. Finally, Notepad is used for typing and loading the files.

The average response time of Citrix XenDesktop and Citrix XenServer for 20 and 30 active users is shown in Figure~\ref{fig5} and Figure~\ref{fig6}, respectively. The Login VSI measures the response time based on six different transactions performed by the simulated users \cite{van2013virtual}. These transactions are Notepad File Open (NFO), Notepad Start Load Document (NSLD), Zip High Compression (ZHC), Zip Low Compression (ZLC), CPU to calculate a large array of random data, and I/O to write the random CPU data array to the disk. The results show that changing the load from medium to heavy will not increase the response time significantly. In addition, the results show that the response time increases as the number of active users increases in both cases. However, the response time does not surpass the threshold value. Note that the calculated threshold values are 2882 and 2998 milliseconds for 20 and 30 active users, respectively. Thus, the system is considered good and no need to add more resources or reduce the allowed number of active users.

\begin{figure}
	\centering           
	\includegraphics[height = 5.20 cm, width=0.49\textwidth]{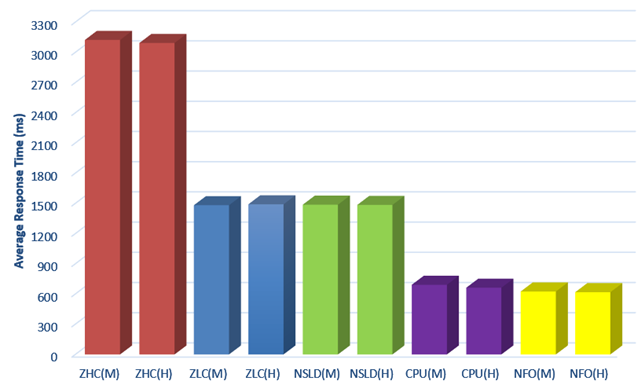}   
	\captionsetup{justification=centering}
	\caption{Average response time for different transactions\\(20 users).} 
	\label{fig7}       
\end{figure}

\begin{figure}
	\centering           
	\includegraphics[height = 5.20 cm, width=0.49\textwidth]{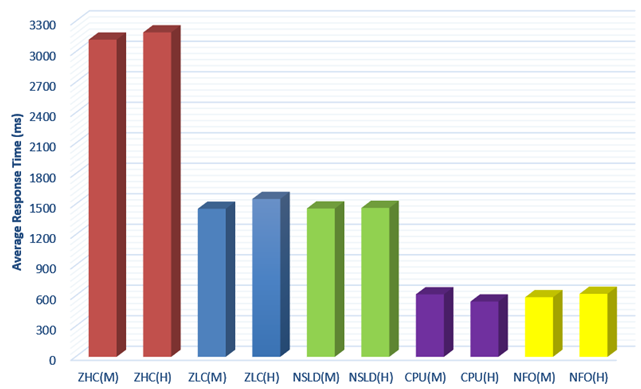}   
	\captionsetup{justification=centering}
	\caption{Average response time for different transactions\\(30 users).} 
	\label{fig8}       
\end{figure}

The average response time for the considered transactions when considering 20 and 30 active users is shown in Figure~\ref{fig7} and Figure~\ref{fig8}, respectively. The results show that the highest response time is recorded for ZHC transaction in both cases. The response time of other transactions are almost the same. In addition, we observed that changing the workload from medium to heavy did not add significant overhead on the system. Thus, the Citrix XenDesktop VDI keeps performing well even when applying heavy workload. Note that the response time of I/O is not mentioned since it is very low, about 8 to 15 milliseconds.

\begin{figure}
	\centering           
	\includegraphics[height = 5.20 cm, width=0.49\textwidth]{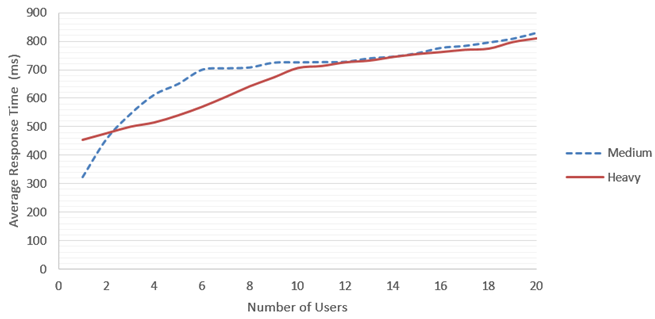}   
	\captionsetup{justification=centering}
	\caption{Average response time for CPU transaction (20 users).} 
	\label{fig9}       
\end{figure}

\begin{figure}
	\centering           
	\includegraphics[height = 5.20 cm, width=0.49\textwidth]{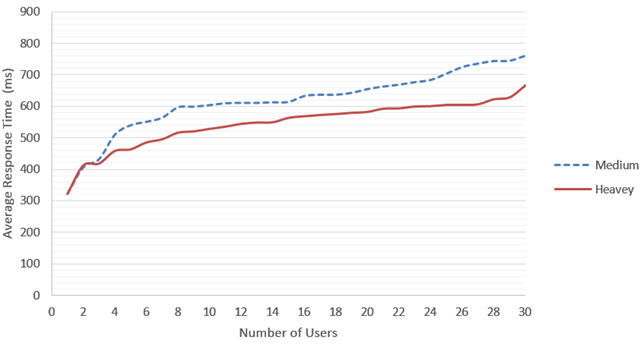}   
	\captionsetup{justification=centering}
	\caption{Average response time for CPU transaction (30 users).} 
	\label{fig10}       
\end{figure}

The average response time of the CPU when considering 20 and 30 active users is shown in Figure~\ref{fig9} and Figure~\ref{fig10}, respectively. The results show that the CPU average response time of medium workload is higher than that of heavy workload. This is because that the medium workload is designed for a VD with one vCPU while the heavy workload is designed for a VD that having more than one vCPU. Note that the VM used for Citrix XenDesktop has 4 vCPUs.

\begin{figure}
	\centering           
	\includegraphics[height = 5.20 cm, width=0.49\textwidth]{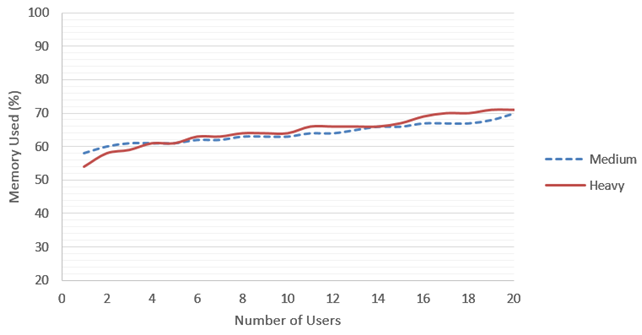}   
	\captionsetup{justification=centering}
	\caption{Memory utilization (20 users).} 
	\label{fig11}       
\end{figure}

\begin{figure}
	\centering           
	\includegraphics[height = 5.20 cm, width=0.49\textwidth]{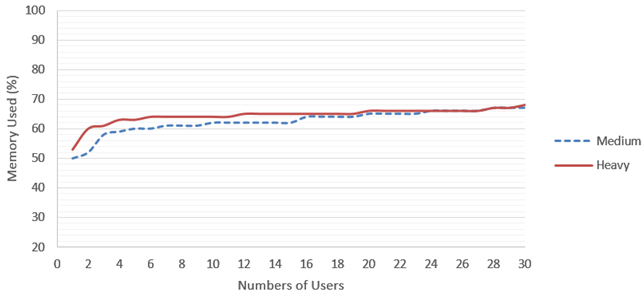}   
	\captionsetup{justification=centering}
	\caption{Memory utilization (30 users).} 
	\label{fig12}       
\end{figure}

The memory utilization for 20 and 30 active users is shown in Figure~\ref{fig11} and Figure~\ref{fig12}, respectively. The results demonstrated that each user has a dedicated memory. Furthermore, the memory utilization is below 70\% for both cases when a number of active users are 20 and 30. That is a good indication of overcoming the paging problem.

\begin{figure}
	\centering           
	\includegraphics[height = 5.2 cm, width=0.49\textwidth]{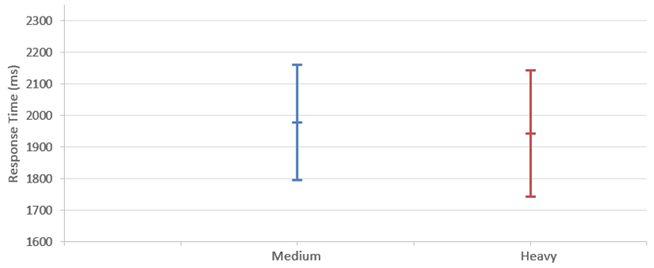}   
	\captionsetup{justification=centering}
	\caption{Confidence Interval for different workloads (20 users).} 
	\label{fig13}       
\end{figure}

\begin{figure}
	\centering           
	\includegraphics[height = 5.2 cm, width=0.49\textwidth]{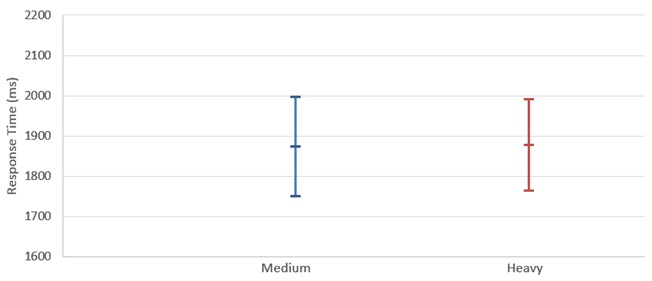}   
	\captionsetup{justification=centering}
	\caption{Confidence Interval for different workloads (20 users).} 
	\label{fig14}       
\end{figure}

\begin{figure}
	\centering           
	\includegraphics[height = 5.2 cm, width=0.49\textwidth]{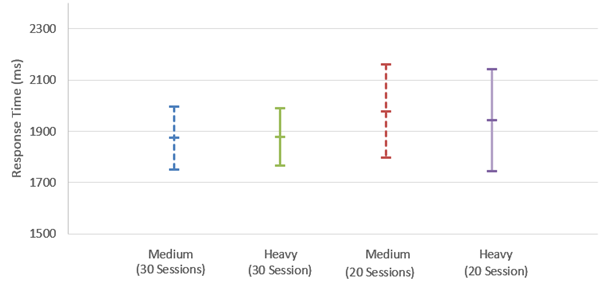}   
	\captionsetup{justification=centering}
	\caption{Confidence Interval for different workloads (20 and 30 users).} 
	\label{fig15}       
\end{figure}

Figure~\ref{fig13}, Figure~\ref{fig14}, and Figure~\ref{fig15} show the Confidence Interval (CI) with considering the significant level as 95\% for both medium and heavy workloads and for 20 and 30 active users. We calculated the CI with maximum error of estimate using Equation~\ref{eq1} \cite{bukh1992art}.

\begin{align}
E =  t_{(n\ -\ 1;\ 1\ -\ \alpha/2)} \cdot \dfrac{s}{\sqrt n}
\label{eq1}
\end{align}

where, $E$ is the maximum error with one degree of confidence ($\alpha$) using two tail distribution (Student $t$ distribution), $s$ is the standard deviation, and $n$ is the number of samples.

It is clear that all CI’s levels are overlapped and the means are also overlapped in each other. This is an indication that the performance of the system did not degrade for both workloads. In addition, increasing the number of active users from 20 to 30 will not affect the average response time. So, we can conclude that the average response time of Citric XenDesktop VDI with Citrix XenServer hypervisor is statistically same even when changing the workload from medium to heavy. Furthermore, the increment in number of active users from 20 to 30 does not affect the overall average response time.

\section{Conclusion and Future Work}

Virtualization technology is used for sharing the abilities of computer systems by splitting the resources among various virtual PCs. Virtual desktop is the best solution for virtualization where every user can use virtual desktop like its own personal PC. In this paper, we successfully build virtual lab for educational organization in which we implemented VDI and created more than 30 virtual desktops on the Citrix XenServer for the students. We measured and evaluated the performance of virtual desktops and Citrix XenServer using Login VSI. The results have shown that the system performs well with 30 virtual desktops without reaching the saturation point. In Login VSI, by gradually increasing the number of simulated users, the system will eventually be saturated. As a result, the performance of the system can be degraded.

Due to the limited available resources, we were able to implement no more than 30 virtual desktops on the Citrix XenServer hypervisor. One of the future directions for this work is to increase the number of virtual desktops and test the performance of the hypervisor in a large scale environment.

Another future direction for this research is the comparison of the obtained results with other VDI environments results. Thus, it will help the organizations in choosing the best VDI platform for their needs.

\section{Acknowledgment}
The authors would like to thank King Fahd University of Petroleum and Minerals (KFUPM) for supporting this research and providing the computing facilities. In addition, the authors would like to acknowledge the support provided by National Science, Technology and Innovation Plan (NSTIP) at KFUPM for funding this work.

\appendices

\ifCLASSOPTIONcaptionsoff
  \newpage
\fi

\bibliography{main}
\bibliographystyle{IEEEtran}

\vfill

\end{document}